\documentclass[pre]{revtex4-2}

\usepackage{amssymb, mathtools, siunitx}
\usepackage{graphicx, color}

\DeclareMathOperator{\erfc}{erfc} 
\DeclareMathOperator{\sgn}{sgn} 

\begin{document}
\title{Deformation of power law in the double Pareto distribution using uniformly distributed observation time}
\author{Ken Yamamoto}
\affiliation{Faculty of Science, University of the Ryukyus, Nishihara, Okinawa 903-0213, Japan}
\author{Takashi Bando, Hirokazu Yanagawa}
\affiliation{Production Engineering Department, SCM Division, Measurement Business Group, ANRITSU CORPORATION, Atsugi, Kanagawa 243-8555, Japan}
\author{Yoshihiro Yamazaki}
\affiliation{School of Advanced Science and Engineering, Waseda University, Shinjuku, Tokyo 169-8555, Japan}

\begin{abstract}
The double Pareto distribution is a heavy-tailed distribution with a power-law tail, that is generated via geometric Brownian motion with an exponentially distributed observation time.
In this study, we examine a modified model wherein the exponential distribution of the observation time is replaced with a continuous uniform distribution.
The probability density, complementary cumulative distribution, and moments of this model are exactly calculated.
Furthermore, the validity of the analytical calculations is discussed in comparison with numerical simulations of stochastic processes.
\end{abstract}

\maketitle

\section{Introduction}
Stochastic models have substantially contributed to the analysis of fluctuating or noisy systems~\cite{vanKampen, Redner}.
Furthermore, simple stochastic processes have been proved to adequately describe real phenomena, and the connection between stochastic processes and resultant probability distributions have been established theoretically.
A probability distribution having a subexponential tail is referred to as a heavy-tailed distribution~\cite{Nair}, which provides theoretical support for statistical physics such as critical phenomena~\cite{Nishimori}, anomalous diffusion~\cite{Avraham}, and long-range memories~\cite{Zwanzig}.
Additionally, heavy-tailed distributions are crucial for complex systems such as social~\cite{Newman, Kobayashi} and biological~\cite{Limpert} systems. 
Lognormal and power-law distributions are the heavy-tailed distributions focused upon in this study.

A random variable $X$ is said to follow the lognormal distribution if the logarithm of $X$ is normally distributed~\cite{Crow}.
The probability density function (PDF) of the lognormal distribution is expressed as
\[
f_\text{LN}(x; \mu,\sigma^2)=\frac{1}{\sqrt{2\pi}\sigma x}\exp\left(-\frac{(\ln x-\mu)^2}{2\sigma^2}\right),
\]
where $\mu$ and $\sigma$ are the mean and standard deviation of $\ln X$, respectively.
Its complementary cumulative distribution function (CCDF) is expressed as
\[
F_\text{LN}(x; \mu,\sigma^2)=\int_x^\infty f_\text{LN}(y; \mu,\sigma^2)dy=\frac{1}{2}\erfc\left(\frac{\ln x-\mu}{\sqrt{2}\sigma}\right),
\]
where
\begin{equation}
\erfc(z)\coloneqq\frac{2}{\sqrt{\pi}}\int_z^\infty e^{-u^2}du
\label{eq:erfc}
\end{equation}
is the complementary error function~\cite{Olver}.
Furthermore, the $k$th moment of the lognormal random variable $X$ is computed as
\begin{equation}
E[X^k]=\exp\left(k\mu+\frac{k^2}{2}\sigma^2\right).
\label{eq:lognormal_moment}
\end{equation}
Specifically, the mean and variance are
\[
E[X] = \exp\left(\mu+\frac{\sigma^2}{2}\right),\quad
V[X] = e^{2\mu+\sigma^2}(e^{\sigma^2}-1),
\]
respectively.

The lognormal distribution occurs in the multiplicative stochastic process, which is given by
\begin{equation}
X_n= M_n X_{n-1},
\label{eq:multiplicative}
\end{equation}
where the initial value $X_0$ is a positive constant, and $M_1, M_2,\ldots$ are random variables distributed independently and identically.
The distribution of $X_n$ for sufficiently large $n$ can be approximated via a lognormal distribution, owing to the central limit theorem.
The process~\eqref{eq:multiplicative} has been used as a simplified model for the X-ray burst~\cite{Uttley} and the growth of organisms~\cite{Yamamoto2016, Koyama}; this model was originally analyzed by Kolmogorov~\cite{Kolmogorov}.

As Eq.~\eqref{eq:multiplicative} has a simple form, various additional effects have been applied.
By modifying Eq.~\eqref{eq:multiplicative}, the lognormal distribution can change qualitatively to other distributions.
For example, power-law distributions are obtained by introducing additive noise~\cite{Takayasu}, reset event~\cite{Manrubia}, random stopping~\cite{Yamamoto2012}, and temporal cumulative sum~\cite{Yamamoto2014, Yamamoto2015}.
The introduction of a lower bound yields the power-law distribution~\cite{Levy}, and a related model introducing the sample-dependent lower bound yields a heavy-tailed but not power-law distribution~\cite{Yamamoto2022}.

Geometric Brownian motion~\cite{Oksendal} can yield a lognormal distribution, and is expressed as the stochastic differential equation
\begin{equation}
dS_t = \mu S_tdt+\sigma S_tdB_t,
\label{eq:GBM}
\end{equation}
where $B_t$ indicates the Brownian motion, $\mu$ is a real constant, and $\sigma$ is a positive constant.
For simplicity, we assume that the initial value $S_0$ is constant.
By using It\^{o}'s lemma~\cite{Oksendal}, the value of $S_t$ at a given time $t=T$ is lognormally distributed, whose PDF is $f_\text{LN}(x;\ln S_0+\tilde\mu T, \sigma^2 T)$, where $\tilde\mu=\mu-\sigma^2/2$.
The geometric Brownian motion is applied to the Black--Scholes equation in mathematical finance~\cite{Paul}.

When the observation time $T$ of the geometric Brownian motion~\eqref{eq:GBM} is changed to an exponential random variable, $S_T$ does not follow the lognormal distribution.
Instead, the double Pareto (DP) distribution is observed~\cite{Reed2004, Mitzenmacher}.
The PDF of $S_T$ can be expressed as
\[
f_\text{DP}(x)=\int_0^\infty \lambda e^{-\lambda t} f_\text{LN}(x;\ln S_0+\tilde\mu t, \sigma^2 t) dt
=\int_0^\infty \frac{\lambda e^{-\lambda t}}{\sqrt{2\pi t}\sigma x}\exp\left(-\frac{(\ln x-\ln S_0-\tilde\mu t)^2}{2\sigma^2t}\right)dt,
\]
where $\lambda e^{-\lambda t}$ is the PDF of the exponential distribution with mean $1/\lambda$.
This integral represents the mixture of the time-dependent lognormal size distribution with the exponential distribution as a weight.

To calculate this integral, $u=\sqrt{t}$ is introduced and the formula~\cite{Olver}
\begin{equation}
\int_0^\infty \exp\left(-a^2u^2-\frac{b^2}{u^2}\right)du=\frac{\sqrt{\pi}}{2a}e^{-2ab}
\label{eq:DP_formula2}
\end{equation}
can be used.
Finally, we obtain
\begin{equation}
f_\text{DP}(x)=
\begin{dcases}
\frac{\lambda}{S_0\sqrt{\tilde{\mu}^2+2\sigma^2\lambda}}\left(\frac{x}{S_0}\right)^{-\alpha-1} & x\ge S_0,\\
\frac{\lambda}{S_0\sqrt{\tilde{\mu}^2+2\sigma^2\lambda}}\left(\frac{x}{S_0}\right)^{\beta-1} & x<S_0,
\end{dcases}
\label{eq:DP_f}
\end{equation}
where
\[
\alpha=\frac{-\tilde\mu+\sqrt{\tilde\mu^2+2\sigma^2\lambda}}{\sigma^2},\quad
\beta=\frac{\tilde\mu+\sqrt{\tilde\mu^2+2\sigma^2\lambda}}{\sigma^2}.
\]
The term ``double Pareto'' originates from the property that $f_\text{DP}(x)$ has two different power-law exponents depending on whether $x\ge S_0$ or $x<S_0$.
The DP distribution has been observed in various phenomena, such as income~\cite{Reed2001} and microblog posting interval~\cite{Wang}.


A natural generalization of the DP distribution involves replacing the exponential distribution of $T$ by other distributions.
Upon replacing $\lambda e^{-\lambda t}$ with a general PDF $g(t)$, the PDF of $S_T$ is formally expressed as
\[
f_g(x)=\int_0^\infty g(t)f_\text{LN}(x;\ln S_0+\tilde\mu t,\sigma^2 t)dt.
\]
However, the calculation of this integral cannot be performed for general $g(t)$.
This study focuses on the case wherein $T$ follows a uniform distribution as a simple case.
In this case, the PDF and CCDF of $S_T$ can be exactly calculated although they have complicated forms.
We further calculate the moments of $S_T$ and compare the CCDF to the discrete-time process~\eqref{eq:multiplicative}.

\section{Probability density}\label{sec2}
In this section, we derive the PDF of the observed value $S_T$ of geometric Brownian motion wherein the observation time $T$ follows the uniform distribution on the interval $[0,T_\text{max}]$:
\[
g_\text{uni}(t)=
\begin{dcases}
\frac{1}{T_\text{max}} & 0\le t\le T_\text{max}\\
0 & t>T_\text{max}.
\end{dcases}
\]

The PDF of $S_T$ in this case is written as $f_\text{uni}(x)$ and is expressed as
\begin{align*}
f_\text{uni}(x)&=\int_0^\infty g_\text{uni}(t)f_\text{LN}(x;\ln S_0+\tilde\mu t,\sigma^2 t)dt\\
&=\frac{1}{T_\text{max}}\int_0^{T_\text{max}} f_\text{LN}(x;\ln S_0+\tilde\mu t,\sigma^2 t)dt\\
&=\frac{1}{T_\text{max}}\int_0^{T_\text{max}}\frac{1}{\sqrt{2\pi t}\sigma x}\exp\left(-\frac{(\ln x-\ln S_0-\tilde\mu t)^2}{2\sigma^2t}\right)dt.
\end{align*}
After changing the integral variable to $u=\sqrt{t/T_\text{max}}$ and some manipulation, we obtain
\[
f_\text{uni}(x)=\frac{2}{\sqrt{2\pi T_\text{max}}\sigma x}\int_0^1\exp\left(-\frac{(\ln x-\ln S_0-T_\text{max}\tilde\mu u^2)^2}{2T_\text{max}\sigma^2 u^2}\right)du.
\]
By introducing the scaling transformations $\mu_\ast=T_\text{max}\mu$, $\sigma_\ast=\sqrt{T_\text{max}}\sigma$, and $\tilde{\mu}_\ast=\mu_\ast-\sigma_\ast^2/2=\tilde{\mu}T_\text{max}$, the parameter $T_\text{max}$ can be eliminated:
\begin{align}
f_\text{uni}(x)&=\frac{2}{\sqrt{2\pi}\sigma_\ast x}\int_0^1\exp\left(-\frac{(\ln x-\ln S_0-\tilde\mu_\ast u^2)^2}{2\sigma_\ast^2 u^2}\right)du\nonumber\\
&=\frac{2}{\sqrt{2\pi}\sigma_\ast x}\left(\frac{x}{S_0}\right)^{\tilde\mu_\ast/\sigma_\ast^2}\int_0^1\exp\left(-\frac{\tilde\mu_\ast^2}{2\sigma_\ast^2}u^2-\frac{(\ln(x/S_0))^2}{2\sigma_\ast^2u^2}\right)du.
\label{eq:uniform_f_scaled}
\end{align}
Therefore, although $f_\text{uni}$ originally involves three parameters $\tilde\mu$, $\sigma$, and $T_\text{max}$, it can be essentially reduced to two ($\tilde\mu_\ast$ and $\sigma_\ast$).

To calculate the integral in Eq.~\eqref{eq:uniform_f_scaled}, we employed the following formula~\cite{Olver}:
\begin{equation}
\int_z^\infty \exp\left(-a^2u^2-\frac{b^2}{u^2}\right)du=\frac{\sqrt{\pi}}{4a}\left[e^{2ab}\erfc\left(az+\frac{b}{z}\right)+e^{-2ab}\erfc\left(az-\frac{b}{z}\right)\right],
\label{eq:uniform_formula}
\end{equation}
where $a,b>0$.
Equation~\eqref{eq:uniform_formula} can be considered as a generalization (or indefinite integral) of Eq.~\eqref{eq:DP_formula2}.
In fact, Eq.~\eqref{eq:DP_formula2} is obtained by considering the $z\to0+$ limit and using the limit values $\erfc(+\infty)=0$ and $\erfc(-\infty)=2$.
Using this relation, we obtain
\begin{align}
\int_0^1\exp\left(-a^2u^2-\frac{b^2}{u^2}\right)du&=\left(\int_0^\infty-\int_1^\infty\right)\exp\left(-a^2u^2-\frac{b^2}{u^2}\right)du
=\frac{\sqrt{\pi}}{4a}\left[e^{-2ab}(2-\erfc(a-b))-e^{2ab}\erfc(a+b)\right]\nonumber\\
&=\frac{\sqrt{\pi}}{4a}\left[e^{-2ab}\erfc(b-a)-e^{2ab}\erfc(a+b)\right],
\label{eq:uniform_formula2}
\end{align}
where $\erfc(-z)=2-\erfc(z)$ (see Ref.~\cite{Olver}) is used in the final equality.

The integral in Eq.~\eqref{eq:uniform_f_scaled} for $\tilde\mu_\ast\ne0$ can be calculated by inserting $a=|\tilde\mu_\ast|/(\sqrt{2}\sigma_\ast)$ and $b=|\ln(x/S_0)|/(\sqrt{2}\sigma_\ast)$ in Eq.~\eqref{eq:uniform_formula2}.
The factor $e^{\pm2ab}$ is reduced to
\[
e^{\pm2ab}=\exp\left(\pm\frac{|\tilde\mu_\ast|}{\sigma_\ast^2}\left|\ln\frac{x}{S_0}\right|\right)
=\left(\frac{x}{S_0}\right)^{\pm\sgn(x-S_0)|\tilde\mu_\ast|/\sigma_\ast^2},
\]
where $\sgn$ is the signum function defined by
\[
\sgn(x)=
\begin{cases}
+1 & x>0,\\
0 & x=0,\\
-1 & x<0.
\end{cases}
\]
Therefore,
\begin{align*}
f_\text{uni}(x)&=\frac{1}{2|\tilde\mu_\ast|x}\left[\left(\frac{x}{S_0}\right)^{(1-\sgn(x-S_0))|\tilde\mu_\ast|/\sigma_\ast^2}\erfc\left(\frac{|\ln(x/S_0)|-|\tilde\mu_\ast|}{\sqrt{2}\sigma_\ast}\right)\right.\\
&\qquad\left.-\left(\frac{x}{S_0}\right)^{(1+\sgn(x-S_0))|\tilde\mu_\ast|/\sigma_\ast^2}\erfc\left(\frac{|\ln(x/S_0)|+|\tilde\mu_\ast|}{\sqrt{2}\sigma_\ast}\right)\right].
\end{align*}
When $\tilde\mu_\ast>0$, the absolute value can be simply removed: $|\tilde\mu_\ast|=\tilde\mu_\ast$.
When $\tilde\mu_\ast<0$, $|\tilde\mu_\ast|=-\tilde\mu_\ast$, and it can be confirmed in a straightforward manner that the expression of $f_\text{uni}(x)$ becomes the same as in $\tilde\mu_\ast>0$.
Thus, we finally obtain
\begin{align}
f_\text{uni}(x)&=\frac{1}{2\tilde\mu_\ast x}\left[\left(\frac{x}{S_0}\right)^{(1-\sgn(x-S_0))\tilde\mu_\ast/\sigma_\ast^2}\erfc\left(\frac{|\ln(x/S_0)|-\tilde\mu_\ast}{\sqrt{2}\sigma_\ast}\right)\right.\nonumber\\
&\qquad\left.-\left(\frac{x}{S_0}\right)^{(1+\sgn(x-S_0))\tilde\mu_\ast/\sigma_\ast^2}\erfc\left(\frac{|\ln(x/S_0)|+\tilde\mu_\ast}{\sqrt{2}\sigma_\ast}\right)\right].
\label{eq:uniform_f}
\end{align}

The $f_\text{uni}(x)$ for $\tilde\mu_\ast=0$ is written as
\[
f_\text{uni}(x)=\frac{2}{\sqrt{2\pi}\sigma_\ast x}\int_0^1\exp\left(-\frac{(\ln(x/S_0))^2}{2\sigma_\ast^2u^2}\right)du
=\frac{2}{\sqrt{2\pi}\sigma_\ast x}\int_0^1\exp\left(-\frac{b^2}{u^2}\right)du,
\]
where $b=\frac{|\ln(x/S_0)|}{\sqrt{2}\sigma_\ast}$, as above.
By introducing a new integration variable $v=|b|/u$ and integrating by parts, we obtain
\begin{align}
f_\text{uni}(x)&=\frac{2}{\sqrt{2\pi}\sigma_\ast x}\left[-\frac{|b|}{v}e^{-v^2}\right]_{|b|}^\infty
-\frac{2}{\sqrt{2\pi}\sigma_\ast x}\int_{|b|}^\infty 2|b|e^{-v^2}dv
=\frac{2e^{-b^2}}{\sqrt{2\pi}\sigma_\ast x}-\frac{\sqrt{2}|b|}{\sigma_\ast x}\erfc(|b|)\nonumber\\
&=\frac{2}{\sqrt{2\pi}\sigma_\ast x}\exp\left(-\frac{(\ln(x/S_0))^2}{2\sigma_\ast^2}\right)-\frac{|\ln(x/S_0)|}{\sigma_\ast^2 x}\erfc\left(\frac{|\ln(x/S_0)|}{\sqrt{2}\sigma_\ast}\right).
\label{eq:uniform_f0}
\end{align}
The derived PDFs in Eqs.~\eqref{eq:uniform_f} and \eqref{eq:uniform_f0} are complicated compared to the DP distribution~\eqref{eq:DP_f}.

We derive the asymptotic form of $f_\text{uni}(x)$ in the $x\to\infty$ and $x\to0+$ limits.
Using the asymptotic expansion of the $\erfc$ function~\cite{Olver}
\begin{equation}
\erfc(z)\sim\frac{1}{\sqrt{\pi}z}e^{-z^2} \quad(z\to\infty)
\label{eq:erfc_asymptotic}
\end{equation}
and the relation
\[
\left(\frac{x}{S_0}\right)^{2\tilde\mu_\ast/\sigma_\ast}\exp\left(-\frac{(\ln(x/S_0)+\tilde\mu_\ast)^2}{2\sigma_\ast^2}\right)=\exp\left(-\frac{(\ln(x/S_0)-\tilde\mu_\ast)^2}{2\sigma_\ast^2}\right),
\]
we obtain
\[
f_\text{uni}(x)\sim\sqrt{\frac{2}{\pi}}\frac{\sigma_\ast}{x(\ln x)^2}\exp\left(-\frac{(\ln(x/S_0)-\tilde\mu_\ast)^2}{2\sigma_\ast^2}\right)
=\frac{2\sigma_\ast^2}{(\ln x)^2}f_\text{LN}(x;\ln S_0+\tilde\mu_\ast, \sigma_\ast^2)\quad (x\to\infty, 0+).
\]
Notably, both limits $x\to\infty$ and $x\to0+$ have the same asymptotic form.
Owing to the $(\ln x)^{-2}$ factor, $f_\text{uni}(x)$ decays slightly faster than the lognormal PDF.

\subsection{Shape of $f_\text{uni}(x)$ graph}\label{subsec2.1}

\begin{figure}[t!]\centering
\raisebox{36mm}{(a)}
\includegraphics[scale=0.8]{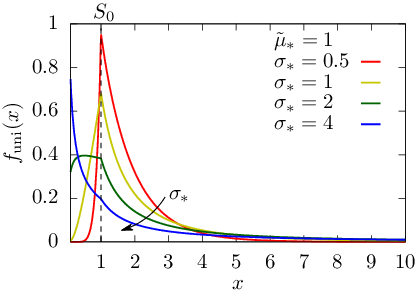}
\hspace{5mm}
\raisebox{36mm}{(b)}
\includegraphics[scale=0.8]{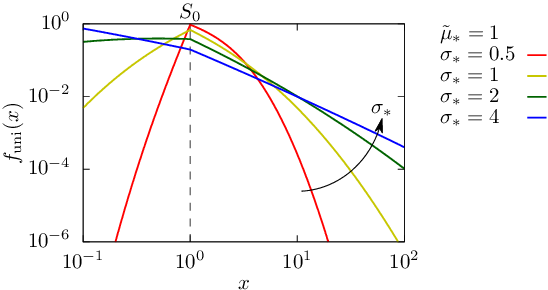}\\[0.5\baselineskip]
\raisebox{36mm}{(c)}
\includegraphics[scale=0.8]{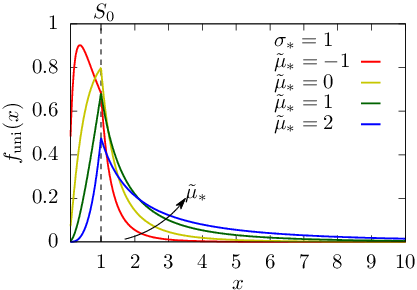}
\hspace{5mm}
\raisebox{36mm}{(d)}
\includegraphics[scale=0.8]{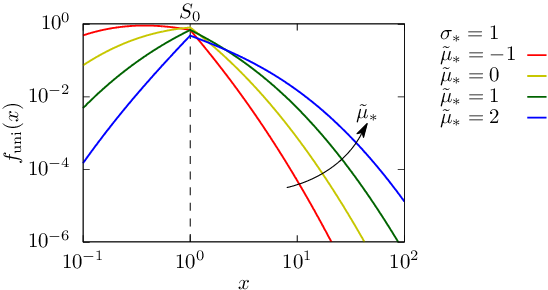}
\caption{
Graphs of $f_\text{uni}(x)$.
(a) $\tilde\mu_\ast=1$ and $\sigma_\ast=0.5,1,2$, and $4$.
(b) Log-log graph of (a).
(c) $\sigma_\ast=1$ and $\tilde\mu_\ast=-1,0,1$, and $2$.
(d) Log-log graph of (c).
The vertical dashed line indicates $x=S_0=1$.
}
\label{fig1}
\end{figure}

The graph of $f_\text{uni}(x)$ is shown in Fig.~\ref{fig1}.
For any $\tilde\mu_\ast$ and $\sigma_\ast$, $f_\text{uni}(x)$ is continuous for all $x>0$ and is not differentiable at $x=S_0$, which corresponds to the discontinuity point of $\sgn(x-S_0)$.
Panel~(a) shows the graph of fixed $\tilde\mu_\ast=1$ and varying $\sigma_\ast$, and (b) shows the log-log graph of (a).
Panel~(c) shows the graph of fixed $\sigma_\ast=1$ and varying $\tilde\mu_\ast$, and (d) shows the log-log graph of (c).
The decay of $f_\text{uni}(x)$ as $x\to\infty$ is slower for larger $\sigma_\ast$ and $\tilde\mu_\ast$.
The graph for $\tilde\mu_\ast=1$ and $\sigma_\ast=4$ in Fig.~\ref{fig1}(a) appears to increase monotonically as $x\to0+$; however, this is not true.
As $\lim_{x\to0+}f_\text{uni}(x)=0$ for any $\tilde\mu_\ast$ and $\sigma_\ast$, this graph attains a maximum value for a very small (but positive) $x$.

In Fig.~\ref{fig1}, the graph of $f_\text{uni}(x)$ has a peak at $x=S_0$ for$(\tilde\mu_\ast, \sigma_\ast)=(1,0.5), (1,1), (0,1)$, and $(2,1)$; the peak for the remaining graphs are not at $x=S_0$.
We investigate the reason for this qualitative change, by deriving the condition for $(\tilde\mu_\ast, \sigma_\ast)$ such that $f_\text{uni}(x)$ has a peak at $x=S_0$.
For sufficiently small $\delta>0$,
\[
f_\text{uni}(S_0\pm\delta)\simeq f_\text{uni}(S_0)\pm f_\text{uni}'(S_0^{\mathord\pm})\delta,
\]
where
\[
f_\text{uni}'(S_0^{\mathord+})=\lim_{h\to 0+}\frac{f_\text{uni}(S_0+h)-f_\text{uni}(S_0)}{h},\quad
f_\text{uni}'(S_0^{\mathord-})=\lim_{h\to 0+}\frac{f_\text{uni}(S_0)-f_\text{uni}(S_0-h)}{h}
\]
are the right and left derivatives of $f_\text{uni}(x)$ at $x=S_0$, respectively.
We have to consider one-sided derivatives because the function $f_\text{uni}(x)$ is not differentiable at $x=S_0$.
The condition such that $f_\text{uni}(x)$ has a peak at $x=S_0$ is
\[
f_\text{uni}'(S_0^{\mathord+})<0,\quad
f_\text{uni}'(S_0^{\mathord-})>0.
\]
The differentiation of Eqs.~\eqref{eq:uniform_f} and \eqref{eq:uniform_f0} yields
\[
f_\text{uni}'(S_0^{\mathord+})=
\begin{dcases}
\frac{1}{\tilde\mu_\ast S_0^2}\left[\left(1-\frac{\tilde\mu_\ast}{\sigma_\ast^2}\right)\erfc\left(\frac{\tilde\mu_\ast}{\sqrt{2}\sigma_\ast}\right)-1\right] & \tilde\mu_\ast\ne0,\\
-\frac{2}{\sqrt{2\pi}\sigma_\ast S_0^2} & \tilde\mu_\ast=0,
\end{dcases}
\]
and
\begin{equation}
f_\text{uni}'(S_0^{\mathord-})=
\begin{dcases}
\frac{1}{\tilde\mu_\ast S_0^2}\left[\frac{2\tilde\mu_\ast}{\sigma_\ast^2}-1+\left(1-\frac{\tilde\mu_\ast}{\sigma_\ast^2}\right)\erfc\left(\frac{\tilde\mu_\ast}{\sqrt{2}\sigma_\ast}\right)\right] & \tilde\mu_\ast\ne0,\\
\frac{1}{\sigma_\ast^2 S_0^2}-\frac{2}{\sqrt{2\pi}\sigma_\ast S_0^2} & \tilde\mu_\ast=0.
\end{dcases}
\label{eq:fprime}
\end{equation}
First, we prove that $f_\text{uni}'(S_0^{\mathord+})<0$ always holds.
This is trivial for $\tilde\mu_\ast=0$; thus, we examine the $\tilde\mu_\ast\ne0$ case:
\[
f_\text{uni}'(S_0^{\mathord+})=-\frac{1}{S_0^2\sigma_\ast^2}\erfc\left(\frac{\tilde\mu_\ast}{\sqrt{2}\sigma_\ast}\right)-\frac{1}{\tilde\mu_\ast S_0^2}\left[1-\erfc\left(\frac{\tilde\mu_\ast}{\sqrt{2}\sigma_\ast}\right)\right].
\]
The first term on the right-hand side is always negative because the $\erfc$ function always takes a positive value.
The factor $1-\erfc(\tilde\mu_\ast/(\sqrt{2}\sigma_\ast))$ in the second term becomes positive for positive $\tilde\mu_\ast$ and becomes negative for negative $\tilde\mu_\ast$.
Thus, $1-\erfc(\tilde\mu_\ast/(\sqrt{2}\sigma_\ast))$ always has the same sign as $\tilde\mu_\ast$, and the second term is always negative.

\begin{figure}[t!]\centering
\includegraphics[scale=0.8]{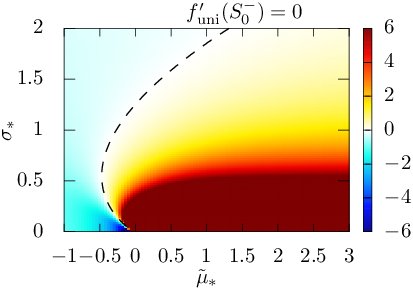}
\caption{
Diagram of $f_\text{uni}'(S_0^{\mathord-})$ on the $(\tilde\mu_\ast, \sigma_\ast)$ plane.
The dashed curve represents the contour of $f_\text{uni}'(S_0^{\mathord-})=0$.
The graph of $f_\text{uni}(x)$ yields a peak at $x=S_0$ if the point $(\tilde\mu_\ast, \sigma_\ast)$ lies on the right side of this curve.
}
\label{fig2}
\end{figure}

In contrast, whether $f_\text{uni}'(S_0^{\mathord-})>0$ depends on $\tilde\mu_\ast$ and $\sigma_\ast$.
Figure~\ref{fig2} shows the $f_\text{uni}'(S_0^{\mathord-})$ values with $S_0=1$ for $-1\le\tilde\mu_\ast\le3$ and $0<\sigma_\ast\le2$.
To ensure clear visualization, we restrict the color bar within the interval $[-6, 6]$, although certain $(\tilde\mu_\ast, \sigma_\ast)$ yield values outside this range [e.g., $f_\text{uni}'(S_0^{\mathord-})\approx49$ for $(\tilde\mu_\ast, \sigma_\ast)=(1,0.2)$].
The dashed curve in the figure represents the contour of $f_\text{uni}'(S_0^{\mathord-})=0$.
Thus, $f_\text{uni}(x)$ exhibits a peak at $x=S_0$ when $(\tilde\mu_\ast, \sigma_\ast)$ is on the right side of this curve.
By using the limit value $\erfc(+\infty)=0$, it can be proven using Eq.~\eqref{eq:fprime} that the curve for $f_\text{uni}'(S_0^{\mathord-})=0$ asymptotically draws the parabola $\tilde\mu_\ast=\sigma_\ast^2/2$ for $\sigma_\ast\gg1$.
When $\tilde\mu_\ast=0$, $f_\text{uni}'(S_0^{\mathord-})>0$ can be exactly solved to attain $\sigma_\ast<\sqrt{\pi/2}\approx1.25$.
However, the solution for the $\tilde\mu_\ast\ne0$ case will not be obtained exactly, owing to the $\erfc$ function.

\begin{figure}[t!]\centering
\includegraphics[scale=0.8]{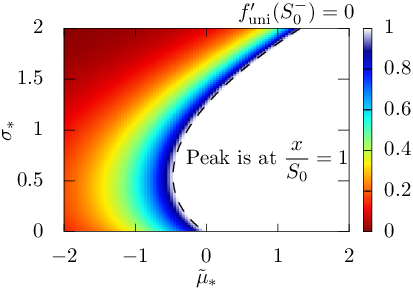}
\caption{
Numerical result for the peak position depicted in the $(\tilde\mu_\ast, \sigma_\ast)$ plane.
The color bar represents the peak position $x$ divided by $S_0$.
The dashed curve shows the contour of $f_\text{uni}'(S_0^{\mathord-})=0$, which is shown in Fig.~\ref{fig2}.
To the right of this curve, $f_\text{uni}(x)$ shows a peak at $x=S_0$.
}
\label{fig3}
\end{figure}

Further analysis shows that $f_\text{uni}(x)$ is unimodal for any $\tilde\mu_\ast$ and $\sigma_\ast$ (further detailes have been presented in Appendix~\ref{appendix}).
Depending on the $(\tilde\mu_\ast, \sigma_\ast)$ values, the peak position of $f_\text{uni}(x)$ is either at $x=S_0$ (corresponding to $f_\text{uni}'(S_0^{\mathord-})\ge0$) or $x<S_0$ ($f_\text{uni}'(S_0^{\mathord-})<0$).
The peak position $x$ for the latter case is characterized by the solution of the equation $f_\text{uni}'(x)=0$.
The explicit form of this equation is presented in Eq.~\eqref{eq:f_solution} in Appendix~\ref{appendix}, which is a transcendental equation involving the $\erfc$ function.
Rather than providing the exact solution of the peak position, we present a numerical result for the peak position in Fig.~\ref{fig3}.
The peak tends to be located at small $x$ with decreasing $\tilde\mu_\ast$ and increasing $\sigma_\ast$.

\section{Complementary cumulative distribution}\label{sec3}
In principle, the CCDF $F_\text{uni}(x)=\int_x^\infty f_\text{uni}(y)dy$ is derived by integrating the PDF $f_\text{uni}(x)$ in Eqs.~\eqref{eq:uniform_f} and \eqref{eq:uniform_f0}.
However, as the absolute value in the $\erfc$ function cannot be handled in a straightforward manner, we first remove the absolute value symbols.

When $x<S_0$ and $\tilde\mu_\ast\ne0$,
\[
f_\text{uni}(x)=\frac{1}{2\tilde\mu_\ast x}\left[\erfc\left(\frac{\ln(x/S_0)-\tilde\mu_\ast}{\sqrt{2}\sigma_\ast}\right)-\left(\frac{x}{S_0}\right)^{2\tilde\mu_\ast/\sigma_\ast^2}\erfc\left(\frac{\ln(x/S_0)+\tilde\mu_\ast}{\sqrt{2}\sigma_\ast}\right)
+2\left(\left(\frac{x}{S_0}\right)^{2\tilde\mu_\ast/\sigma_\ast^2}-1\right)\right],
\]
where the relation $\erfc(-z)=2-\erfc(z)$ is used.
The difference from the case $x>S_0$ is the last term.
Therefore, for any $x>0$,
\[
f_\text{uni}(x)=\frac{1}{2\tilde\mu_\ast x}\left[\erfc\left(\frac{\ln(x/S_0)-\tilde\mu_\ast}{\sqrt{2}\sigma_\ast}\right)-\left(\frac{x}{S_0}\right)^{2\tilde\mu_\ast/\sigma_\ast^2}\erfc\left(\frac{\ln(x/S_0)+\tilde\mu_\ast}{\sqrt{2}\sigma_\ast}\right)
+(1-\sgn(x-S_0))\left(\left(\frac{x}{S_0}\right)^{2\tilde\mu_\ast/\sigma_\ast^2}-1\right)\right]
\]
is valid.

For the derivation of $F_\text{uni}(x)$, we calculate the following two integrals as a preliminary step:
\[
I_1=\int_x^\infty \erfc\left(\frac{\ln(y/S_0)-\tilde\mu_\ast}{\sqrt{2}\sigma_\ast}\right)\frac{dy}{y},
\quad
I_2=\int_x^\infty \left(\frac{y}{S_0}\right)^{2\tilde\mu_\ast/\sigma_\ast^2}\erfc\left(\frac{\ln(y/S_0)+\tilde\mu_\ast}{\sqrt{2}\sigma_\ast}\right)\frac{dy}{y}.
\]
By setting $z=(\ln(y/S_0)-\tilde\mu_\ast)/(\sqrt{2}\sigma_\ast)$ and $z_{\mathord-}=(\ln(x/S_0)-\tilde\mu_\ast)/(\sqrt{2}\sigma_\ast)$,
\begin{align*}
I_1&=\sqrt{2}\sigma_\ast\int_{z_{\mathord-}}^\infty \erfc(z)dz
=\left[\sqrt{2}\sigma_\ast z\erfc(z)\right]_{z_{\mathord-}}^\infty+2\sqrt{\frac{2}{\pi}}\sigma_\ast\int_{z_{\mathord-}}^\infty ze^{-z^2}dz\\
&=-\left(\ln\frac{x}{S_0}-\tilde\mu_\ast\right)\erfc\left(\frac{\ln(x/S_0)-\tilde\mu_\ast}{\sqrt{2}\sigma_\ast}\right)+\sqrt{\frac{2}{\pi}}\sigma_\ast\exp\left(-\frac{(\ln(x/S_0)-\tilde\mu_\ast)^2}{2\sigma_\ast^2}\right).
\end{align*}
We use integration by parts in the second equality.
Similarly, by setting $z=(\ln(y/S_0)+\tilde\mu_\ast)/(\sqrt{2}\sigma_\ast)$ and $z_{\mathord+}=(\ln(x/S_0)+\tilde\mu_\ast)/(\sqrt{2}\sigma_\ast)$,
\begin{align*}
I_2&=\sqrt{2}\sigma_\ast e^{-2\tilde\mu_\ast^2/\sigma_\ast^2}\int_{z_{\mathord+}}^\infty \exp\left(\frac{2\sqrt{2}\tilde\mu_\ast}{\sigma_\ast}z\right)\erfc(z)dz\\
&=\left[\frac{\sigma_\ast^2}{2\tilde\mu_\ast}\exp\left(-\frac{2\tilde\mu_\ast^2}{\sigma_\ast^2}+\frac{2\sqrt{2}\tilde\mu_\ast}{\sigma_\ast}z\right)\erfc(z)\right]_{z_{\mathord+}}^\infty
+\frac{\sigma_\ast^2}{\tilde\mu_\ast\sqrt{\pi}}e^{-2\tilde\mu_\ast^2/\sigma_\ast^2}\int_{z_{\mathord+}}^\infty\exp\left(-z^2+\frac{2\sqrt{2}\tilde\mu_\ast}{\sigma_\ast}z\right)dz\\
&=-\frac{\sigma_\ast^2}{2\tilde\mu_\ast}\left(\frac{x}{S_0}\right)^{2\tilde\mu_\ast/\sigma_\ast^2}\erfc\left(\frac{\ln(x/S_0)+\tilde\mu_\ast}{\sqrt{2}\sigma_\ast}\right)+\frac{\sigma_\ast^2}{\sqrt{\pi}\tilde\mu_\ast}\int_{z_{\mathord+}-\sqrt{2}\tilde\mu_\ast/\sigma_\ast}^\infty e^{-z^2}dz\\
&=-\frac{\sigma_\ast^2}{2\tilde\mu_\ast}\left(\frac{x}{S_0}\right)^{2\tilde\mu_\ast/\sigma_\ast^2}\erfc\left(\frac{\ln(x/S_0)+\tilde\mu_\ast}{\sqrt{2}\sigma_\ast}\right)+\frac{\sigma_\ast^2}{2\tilde\mu_\ast}\erfc\left(\frac{\ln(x/S_0)-\tilde\mu_\ast}{\sqrt{2}\sigma_\ast}\right).
\end{align*}

The CCDF for $x\ge S_0$ becomes
\begin{align*}
F_\text{uni}(x)&=\frac{1}{2\tilde\mu_\ast}\int_x^\infty \left[\erfc\left(\frac{\ln(y/S_0)-\tilde\mu_\ast}{\sqrt{2}\sigma_\ast}\right)-\left(\frac{y}{S_0}\right)^{2\tilde\mu_\ast/\sigma_\ast^2}\erfc\left(\frac{\ln(y/S_0)+\tilde\mu_\ast}{\sqrt{2}\sigma_\ast}\right)\right]\frac{dy}{y}\\
&=\frac{I_1-I_2}{2\tilde\mu_\ast}\\
&=\frac{1}{2\tilde\mu_\ast}\left[\sqrt{\frac{2}{\pi}}\sigma_\ast\exp\left(-\frac{(\ln(x/S_0)-\tilde\mu_\ast)^2}{2\sigma_\ast^2}\right)+\frac{\sigma_\ast^2}{2\tilde\mu_\ast}\left(\frac{x}{S_0}\right)^{2\tilde\mu_\ast/\sigma_\ast^2}\erfc\left(\frac{\ln(x/S_0)+\tilde\mu_\ast}{\sqrt{2}\sigma_\ast}\right)\right.\\
&\qquad\left.-\left(\ln\frac{x}{S_0}-\tilde\mu_\ast+\frac{\sigma_\ast^2}{2\tilde\mu_\ast}\right)\erfc\left(\frac{\ln(x/S_0)-\tilde\mu_\ast}{\sqrt{2}\sigma_\ast}\right)\right],
\end{align*}
and for $x<S_0$,
\[
F_\text{uni}(x)=\frac{I_1-I_2}{2\tilde\mu_\ast}+\frac{1}{\tilde\mu_\ast}\int_x^{S_0} \left(\left(\frac{y}{S_0}\right)^{2\tilde\mu_\ast/\sigma_\ast^2}-1\right)\frac{dy}{y}.
\]
The remaining integral can be calculated as
\[
\frac{1}{\tilde\mu_\ast}\int_x^{S_0} \left(\left(\frac{y}{S_0}\right)^{2\tilde\mu_\ast/\sigma_\ast^2}-1\right)\frac{dy}{y}
=\frac{1}{\tilde\mu_\ast}\left(\frac{\sigma_\ast^2}{2\tilde\mu_\ast}-\frac{\sigma_\ast^2}{2\tilde\mu_\ast}\left(\frac{x}{S_0}\right)^{2\tilde\mu_\ast/\sigma_\ast^2}+\ln\frac{x}{S_0}\right).
\]

Consequently, we obtain
\begin{align}
F_\text{uni}(x)&=\frac{1}{2\tilde\mu_\ast}\left[\sqrt{\frac{2}{\pi}}\sigma_\ast\exp\left(-\frac{(\ln(x/S_0)-\tilde\mu_\ast)^2}{2\sigma_\ast^2}\right)+\frac{\sigma_\ast^2}{2\tilde\mu_\ast}\left(\frac{x}{S_0}\right)^{2\tilde\mu_\ast/\sigma_\ast^2}\erfc\left(\frac{\ln(x/S_0)+\tilde\mu_\ast}{\sqrt{2}\sigma_\ast}\right)\right.\nonumber\\
&\qquad\left.-\left(\ln\frac{x}{S_0}-\tilde\mu_\ast+\frac{\sigma_\ast^2}{2\tilde\mu_\ast}\right)\erfc\left(\frac{\ln(x/S_0)-\tilde\mu_\ast}{\sqrt{2}\sigma_\ast}\right)\right]\nonumber\\
&\quad+\frac{1-\sgn(x-S_0)}{2\tilde\mu_\ast}\left(\frac{\sigma_\ast^2}{2\tilde\mu_\ast}-\frac{\sigma_\ast^2}{2\tilde\mu_\ast}\left(\frac{x}{S_0}\right)^{2\tilde\mu_\ast/\sigma_\ast^2}+\ln\frac{x}{S_0}\right),
\label{eq:uniform_F}
\end{align}
which is valid for all $x>0$.
This result is extremely complex, but it is exact.

By integrating Eq.~\eqref{eq:uniform_f0} or taking the $\tilde\mu_\ast\to0$ limit in Eq.~\eqref{eq:uniform_F}, $F_\text{uni}(x)$ for $\tilde\mu_\ast=0$ is obtained as
\[
F_\text{uni}(x)=\frac{1}{2}\erfc\left(\frac{\ln(x/S_0)}{\sqrt{2}\sigma_\ast}\right)-\frac{1}{2\sqrt{2\pi}\sigma_\ast}\exp\left(-\frac{(\ln(x/S_0))^2}{2\sigma_\ast^2}\right)+\frac{1}{2\sigma_\ast^2}\left(\ln\frac{x}{S_0}\right)^2\erfc\left(\frac{\ln(x/S_0)}{\sqrt{2}\sigma_\ast}\right).
\]

\begin{figure}[t!]\centering
\raisebox{36mm}{(a)}
\includegraphics[scale=0.8]{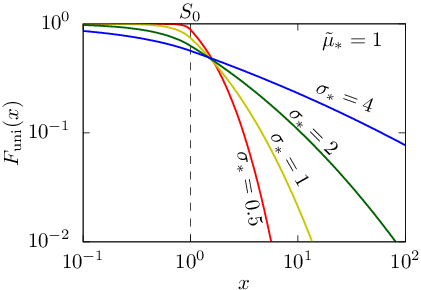}
\hspace{5mm}
\raisebox{36mm}{(b)}
\includegraphics[scale=0.8]{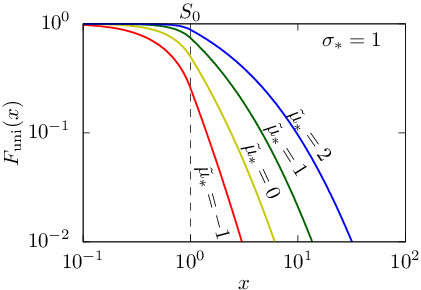}
\caption{
Log-log graphs of $F_\text{uni}(x)$ for $\tilde\mu_\ast=1$ and $\sigma_\ast=0.5,1,2$, and $4$ (a), and for $\sigma_\ast=1$ and $\tilde\mu_\ast=-1,0,1$, and $2$ (b).
The vertical dashed line indicates $x=S_0=1$.
}
\label{fig4}
\end{figure}

Figure~\ref{fig4} shows the graph of $F_\text{uni}(x)$ on a log-log scale.
The graph decays slower for larger $\sigma_\ast$ [in (a)] and larger $\tilde\mu_\ast$ [in (b)].
The graphs do not exhibit significant variety in shape as $f_\text{uni}(x)$ in Fig.~\ref{fig2}, and the overall shapes are similar for different $\tilde\mu_\ast$ and $\sigma_\ast$,

\section{Calculation of moments}
The $k$th moment of $S_T$ can be calculated as
\begin{align*}
E[S_T^k]&=\int_0^\infty x^k f_\text{uni}(x)dx\\
&=\int_0^\infty x^k \frac{1}{T_\text{max}}\int_0^{T_\text{max}} f_\text{LN}(x;\ln S_0+\tilde\mu t,\sigma^2 t)dt dx\\
&=\frac{1}{T_\text{max}}\int_0^{T_\text{max}} \int_0^\infty x^k f_\text{LN}(x;\ln S_0+\tilde\mu t,\sigma^2 t)dx dt\\
&=\frac{S_0^k}{T_\text{max}}\int_0^{T_\text{max}}\exp\left(k\tilde\mu t+\frac{k^2}{2}\sigma^2 t\right)dt\\
&=\frac{2}{k^2\sigma_\ast^2+2k\tilde\mu_\ast}\left[\exp\left(\frac{k^2\sigma_\ast^2}{2}+k\tilde\mu_\ast\right)-1\right]S_0^k.
\end{align*}
The moment of the lognormal distribution [Eq.~\eqref{eq:lognormal_moment}] is used in the fourth equality.
In contrast to the DP distribution, $E[S_T^k]$ is not divergent for any $k$.

Specifically, the mean and variance of $S_T$ for $\mu_\ast\ne0$ and $\mu_\ast\ne-\sigma_\ast^2$ becomes
\[
E[S_T]=\frac{e^{\mu_\ast}-1}{\mu_\ast}S_0,\quad
V[S_T]=\frac{e^{\sigma_\ast^2+\mu_\ast}-1}{\sigma_\ast^2+\mu_\ast}S_0^2-\frac{(e^{\mu_\ast}-1)^2}{\mu_\ast^2}S_0^2.
\]
We obtain $E[S_T]=S_0$ when $\mu_\ast=0$ and
\[
V[S_T]=S_0^2-\frac{(e^{\mu_\ast}-1)^2}{\mu_\ast^2}S_0^2
\]
when $\mu_\ast=-\sigma_\ast^2$.

\begin{figure}[t!]\centering
\raisebox{36mm}{(a)}
\includegraphics[scale=0.8]{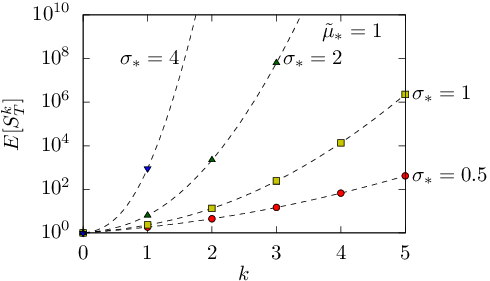}
\hspace{3mm}
\raisebox{36mm}{(b)}
\includegraphics[scale=0.8]{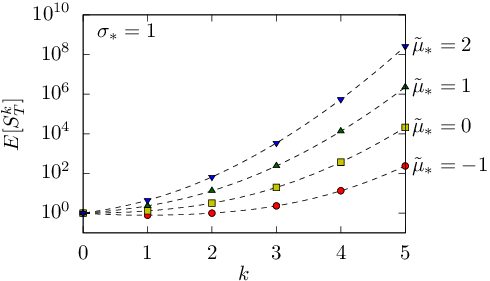}
\caption{
The $k$th moment of $S_T$ ($S_0=1$) up to $k=5$ for $\tilde\mu_\ast=1$ and $\sigma_\ast=0.5, 1, 2$, and $4$ (a),
and for $\sigma_\ast=1$ and $\tilde\mu_\ast=-1,0,1$, and $2$ (b).
}
\label{fig5}
\end{figure}

Figure~\ref{fig5} shows the semi-log graphs of $E[S_T^k]$ for $S_0=1$ with different $\tilde\mu_\ast$ and $\sigma_\ast$.
$E[S_T^k]$ becomes large for large $\tilde\mu_\ast$ and $\sigma_\ast$.
The graph for $\tilde\mu_\ast=-1$ and $\sigma_\ast=1$ in Fig.~\ref{fig5}(b) is not monotonically increasing; $E[S_T^1]\approx0.787<1=E[S_T^0]$.
We can easily prove that
\[
\frac{2}{k^2\sigma_\ast^2+2k\tilde\mu_\ast}\left[\exp\left(\frac{k^2\sigma_\ast^2}{2}+k\tilde\mu_\ast\right)-1\right]<1
\]
when
\[
\frac{k^2\sigma_\ast^2}{2}+k\tilde\mu_\ast<0.
\]
Therefore, this phenomenon can occur for the $\tilde\mu_\ast<0$ case.

\section{Comparison to discrete-time multiplicative process}\label{sec5}
We investigate whether the above results are valid for the discrete-time multiplicative stochastic process~\eqref{eq:multiplicative}.

If $X_0$ is a constant value, $\ln X_n=\ln M_1+\cdots+\ln M_n +\ln X_0$ approximately follows the normal distribution with mean $\mu n$ and variance $\sigma^2 n$, where $\mu=E[\ln M_i]$ and $\sigma^2=V[\ln M_i]$.
Therefore, $X_n$ approximately follows the lognormal distribution with $f_\text{LN}(x;\ln X_0+\mu n, \sigma^2n)$.
This lognormality is similar to $S_T$ for the geometric Brownian motion having $f_\text{LN}(x;\ln S_0+\tilde\mu T, \sigma^2T)$.

If the observation time $n$ is drawn from the discrete uniform distribution on $\{1,2,\ldots,N\}$, the PDF and CCDF of $X_n$ are expected to become approximately similar to $f_\text{uni}(x)$ and $F_\text{uni}(x)$ derived in Sections~\ref{sec2} and \ref{sec3}, respectively.
We compare the numerically generated CCDF and $F_\text{uni}(x)$ using two examples for the distribution of $M_i$.

\begin{figure}[t!]\centering
\raisebox{36mm}{(a)}
\includegraphics[scale=0.8]{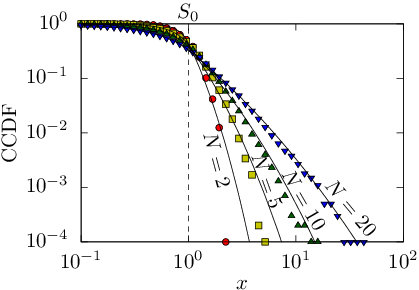}
\hspace{5mm}
\raisebox{36mm}{(b)}
\includegraphics[scale=0.8]{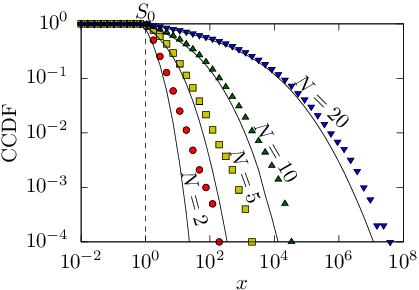}
\caption{
Comparison between numerically calculated CCDF of $X_n$ for $N=2,5,10$, and $20$ (points) and the corresponding $F_\text{uni}(x)$ (curves).
The random variable $M_i$ is drawn from the uniform distribution on the interval$[1/2, 3/2]$ in (a) and the power-law distribution with exponent $\nu=3$ and lower bound $m_0=1$ in (b).
}
\label{fig6}
\end{figure}

The first example is the case wherein $M_i$ is uniformly distributed over the interval $[a,b]$ ($a>0$).
The mean $\mu$ and variance $\sigma^2$ of $\ln M_i$ become
\[
\mu=\int_a^b \frac{1}{b-a} \ln y dy = \frac{b\ln b-a\ln a}{b-a}-1,\quad
\sigma^2=\int_a^b \frac{1}{b-a}(\ln y-\mu)^2dy = 1-ab\left(\frac{\ln b-\ln a}{b-a}\right)^2.
\]
In the numerical calculation, we set $a=1/2$ and $b=3/2$ so that $\mu\approx-0.0452$ and $\sigma^2\approx0.0948$.
The CCDFs for $N=2,5,10$, and $20$ are shown in Fig.~\ref{fig6}(a) with points, calculated from $10^4$ independent samples each.
The solid curves represent $F_\text{uni}(x)$ in Eq.~\eqref{eq:uniform_F} by using $\tilde\mu_\ast=\mu N$ and $\sigma_\ast^2=\sigma^2N$.

The second example is where $M_i$ follows the power-law distribution.
The mean $\mu$ and variance $\sigma^2$ of $\ln M_i$ can be computed to be
\[
\mu=\int_{m_0}^\infty \frac{\nu-1}{m_0}\left(\frac{y}{m_0}\right)^{-\nu}\ln y dy=\frac{1}{\nu-1}+\ln m_0,\quad
\sigma^2=\int_{m_0}^\infty \frac{\nu-1}{m_0}\left(\frac{y}{m_0}\right)^{-\nu}(\ln y-\mu)^2 dy=\frac{1}{(\nu-1)^2},
\]
respectively, where the power-law PDF with exponent $\nu(>1)$ and lower bound $m_0(>0)$ is expressed as
\[
\frac{\nu-1}{m_0}\left(\frac{y}{m_0}\right)^{-\nu}.
\]
In the numerical calculation, we used  $\nu=3$ and $m_0=1$, so that $\mu=1/2$ and $\sigma^2=1/4$.
Figure~\ref{fig6}(b) shows the numerical result for $N=2,5,10$, and $20$ with points and $F_\text{uni}(x)$ with curves.

A common property in these two examples is that the CCDF of $X_n$ gradually approaches $F_\text{uni}(x)$ with increase in $N$.
The deviation from $F_\text{uni}(x)$ in small $N$ is attributed to the fact that the lognormal approximation based on the central limit theorem fails.
In other words, the distribution of $X_n$ for small $N$ is dependent on the individuality of the distribution of $M_i$.
For example, the numerical CCDF of $X_n$ for the uniform $M_i$ decays faster than $F_\text{uni}(x)$, whereas that for the power-law $M_i$ decays slower than $F_\text{uni}(x)$.
This observation is consistent with the property that the lognormal distribution has a tail heavier than the uniform distribution and the power-law distribution has an even further heavier tail.

\section{Discussion}
This study examined the geometric Brownian motion $S_T$ under uniformly distributed observation time $T$.
This model is a deformation of the DP distribution in which the observation time follows an exponential distribution.
The PDF $f_\text{uni}(x)$ and CCDF $F_\text{uni}(x)$ are exactly calculated, although they have nontrivial complicated forms compared to those of the DP distribution.
Consequently, the power-law form of the DP distribution is realized owing to the balance of the geometric Brownian motion and the exponential distribution of $T$.

By changing the observation time distribution, the DP distribution can be generalized.
However, it remains unclear whether the exact expressions of the PDF and CCDF can be obtained in such a generalized case.
In particular, the difficulty will arise in the computation of the (indefinite) integral, which involves $\exp(-a^2u^2-b^2/u^2)$ factor.
Therefore, observation time distributions that provide exact PDF and CCDF of $S_T$ will be highly limited.

In this study, we assume that the initial value of the geometric Brownian motion, $S_0$, is a constant, which is common to the DP distribution~\cite{Mitzenmacher}.
A possible extension of this study involves changing $S_0$ to a random variable.
When $S_0$ is distributed lognormally and the observation time $T$ is exponentially distributed, the distribution of $S_T$ is exactly calculated and is referred to as the double Pareto-lognormal distribution~\cite{Reed2004, Grbac}.
In the context of the double Pareto-lognormal distribution, an interesting challenge associated with this study is the derivation of the distribution of $S_T$ with uniform $T$ when the constant $S_0$ in this study is replaced with a lognormal random variable.

In Section~\ref{sec5}, the applicability of $F_\text{uni}(x)$ to discrete-time stochastic process~\eqref{eq:multiplicative} is examined.
$F_\text{uni}(x)$ is expected to provide a reasonable estimate when the maximum number $N$ of steps is large, whereas the distribution of $X_n$ is strongly dependent on the property of the random variable $M_i$ for small $N$.
A further systematic and quantitative study is required to establish the connection between continuous- and discrete-time processes.

From a practical viewpoint, the exploration of empirical datasets that exhibit the proposed distribution is an important future research direction.
Because the double Pareto distribution has been observed in various fields related to human activities and social phenomena~\cite{Reed2001, Wang}, we believe that the proposed distribution, which can be regarded as a modification of the DP distribution, is useful in the analysis of empirical data.
In future studies, the practical importance of this study will be tested by its application to data analysis.

\section{Conclusion}
The double Pareto (DP) distribution, a double-sided power-law distribution, can be obtained by geometric Brownian motion with a constant initial value and exponentially distributed observation time.
This study investigates a deformation of the DP distribution by replacing the exponential distribution for the observation time with a continuous uniform distribution.

The exact forms of the probability density function (PDF) and complementary cumulative distribution function (CCDF) are derived using the error function [see Eqs.~\eqref{eq:uniform_f} and \eqref{eq:uniform_F}].
Furthermore, the detailed shape of the PDF, e.g., the asymptotic form, unimodality, and peak position, is analyzed.
%
The main finding of this study is the establishment of a parametric distribution that is an extension of the lognormal and power-law distributions.

\appendix
\section{Unimodality of $f_\text{uni}(x)$}\label{appendix}
Here we show that the PDF $f_\text{uni}(x)$ given in Eqs.~\eqref{eq:uniform_f} and \eqref{eq:uniform_f0} is unimodal.
That is, $f_\text{uni}(x)$ has exactly one peak, at $x=S_0$ or $0<x<S_0$.

\begin{figure}[t!]\centering
\raisebox{36mm}{(a)}\hspace{-3mm}
\includegraphics[scale=0.8]{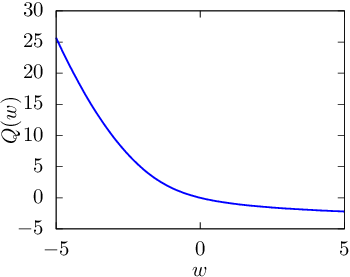}
\hspace{5mm}
\raisebox{36mm}{(b)}\hspace{-3mm}
\includegraphics[scale=0.8]{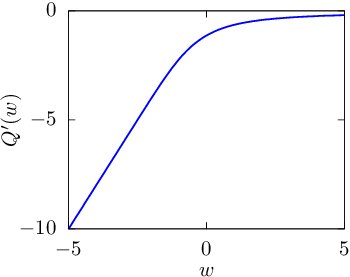}
\hspace{5mm}
\raisebox{36mm}{(c)}\hspace{-3mm}
\includegraphics[scale=0.8]{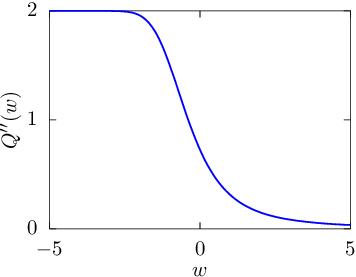}
\caption{
Graphs of $Q(w)$, $Q'(w)$, and $Q''(w)$ in $-5\le w\le5$.
(a) The function $Q(w)$ is strictly convex.
(b) The derivative $Q'(w)$ is increasing and negative-valued.
(c) The second derivative $Q''(w)$ is positive-valued.
}
\label{fig7}
\end{figure}

Before the main proof of the unimodality, we prove that the function
\[
Q(w)=\ln[e^{w^2}\erfc(w)]
\]
is strictly convex.
The second derivative of $Q(w)$ becomes
\[
Q''(w)=\frac{2e^{-w^2}}{\erfc(w)^2}\left[e^{w^2}\erfc(w)^2+\frac{2}{\sqrt{\pi}}w\erfc(w)-\frac{2}{\pi}e^{-w^2}\right]
\eqqcolon\frac{2e^{-w^2}}{\erfc(w)^2}q(w).
\]
Notably, the key to a simple proof is to separate $e^{-w^2}$ in the numerator.
We analyze the function $q(w)$.
The derivative of $q(w)$ becomes
\[
q'(w)=2e^{w^2}\erfc(w)\left[w\erfc(w)-\frac{e^{-w^2}}{\sqrt{\pi}}\right]=-2e^{w^2}\erfc(w)\int_{w}^\infty \erfc(u)du,
\]
where
\[
\int_w^\infty \erfc(u)du=-w\erfc(w)+\frac{e^{-w^2}}{\sqrt{\pi}}
\]
can be derived by integration by parts.
The $\erfc$ function always takes a positive value, and its integral always becomes positive.
Therefore, $q'(w)<0$, i.e., $q(w)$ is a decreasing function.
Using the asymptotic expansion~\eqref{eq:erfc_asymptotic} of the $\erfc$ function, we obtain
\[
\lim_{w\to\infty}q(w)=0.
\]
Consequently, $q(w)$ takes a positive value for any $w$.
Thus, $Q''(w)>0$, i.e., the strict convexity of $Q(w)$, is proven.
For reference, Fig.~\ref{fig7} shows graphs of $Q(w)$, $Q'(w)$, and $Q''(w)$.

If $f_\text{uni}(x)$ has a peak at $x\ne S_0$, this peak position is characterized by the equation $f_\text{uni}'(x)=0$.
First, we prove the unimodality for the $\tilde\mu_\ast=0$ case.
By differentiating Eq.~\eqref{eq:uniform_f0}, the equation $f_\text{uni}'(x)=0$ becomes
\[
\left(1-\ln\frac{x}{S_0}\right)\erfc\left(\pm\frac{\ln(x/S_0)}{\sqrt{2}\sigma_\ast}\right)=\mp\sqrt{\frac{2}{\pi}}\sigma_\ast\exp\left(-\frac{(\ln(x/S_0))^2}{2\sigma_\ast^2}\right),
\]
where the upper and lower signs refer to the cases of $x>S_0$ and $x<S_0$, respectively.
Taking the logarithm of this equation for $x<S_0$ and introducing $w=-\ln(x/S_0)/(\sqrt{2}\sigma_\ast)$, we obtain
\begin{equation}
Q'(w)=-\frac{\sqrt{2}}{\sigma_\ast},
\label{eq:Q_0}
\end{equation}
where
\[
Q'(w)=-\frac{2e^{-w^2}}{\sqrt{\pi}\erfc(w)}+2w.
\]
Owing to the strict convexity of $Q(w)$, the derivative $Q'(w)$ is an increasing function.
Moreover, $Q'(w)$ satisfies
\[
\lim_{w\to-\infty}Q'(w)=-\infty,\quad
\lim_{w\to\infty}Q'(w)=0,
\]
where the asymptotic expansion~\eqref{eq:erfc_asymptotic} is required to compute $Q'(\infty)$.
Therefore, Eq.~\eqref{eq:Q_0} has a unique solution $w$ for any $\sigma_\ast>0$. 
If this solution is $w>0$, the peak position of $f_\text{uni}(x)$ is $x=S_0\exp(-\sqrt{2}\sigma_\ast w)<S_0$, and otherwise, $f_\text{uni}(x)$ is monotonically increasing in $0<x<S_0$ and the peak is at $x=S_0$.
The solution $w$ is positive if and only if
\[
-\frac{\sqrt{2}}{\sigma_\ast}<Q'(0)=-\frac{2}{\sqrt{\pi}},
\]
which is equivalent to
\[
\sigma_\ast>\sqrt{\frac{\pi}{2}}.
\]
This threshold value $\sqrt{\pi/2}$ can be obtained as the solution of $f_\text{uni}'(S_0^{-})=0$ stated in Section~\ref{subsec2.1}.

The equation for $x>S_0$ becomes
\[
Q'(w)=\frac{\sqrt{2}}{\sigma_\ast},
\]
with $w=\ln(x/S_0)/(\sqrt{2}\sigma_\ast)$.
This equation does not have solutions because $Q'(w)$ is always negative, as stated above; $f_\text{uni}(x)$ does not have a peak in $x>S_0$.
Thus, the unimodality of $f_\text{uni}(x)$ for $\tilde\mu_\ast=0$ is proven.

Next, we show the unimodality of $f_\text{uni}(x)$ for $\tilde\mu_\ast\ne0$.
The equation $f_\text{uni}'(x)=0$ becomes
\begin{equation}
\erfc\left(\pm\frac{\ln(x/S_0)-\tilde\mu_\ast}{\sqrt{2}\sigma_\ast}\right)
=\left(1-\frac{2\tilde\mu_\ast}{\sigma_\ast^2}\right)\left(\frac{x}{S_0}\right)^{2\tilde\mu_\ast/\sigma_\ast^2}\erfc\left(\pm\frac{\ln(x/S_0)+\tilde\mu_\ast}{\sqrt{2}\sigma_\ast}\right),
\label{eq:f_solution}
\end{equation}
where $\pm$ indicates ``$+$'' for $x>S_0$ and ``$-$'' for $x<S_0$, as in the $\tilde\mu_\ast=0$ case above.
Since the $\erfc$ function is positive-valued, the existence of a solution $x$ requires $1-2\tilde\mu_\ast/\sigma_\ast^2>0$.
Otherwise, if $1-2\tilde\mu_\ast/\sigma_\ast^2\le0$, Eq.~\eqref{eq:f_solution} does not have solutions, which means that the peak of $f_\text{uni}(x)$ is $x=S_0$.
Taking the logarithm of Eq.~\eqref{eq:f_solution} and introducing $w=\pm\ln(x/S_0)/(\sqrt{2}\sigma_\ast)$, we obtain
\begin{equation}
Q\left(w+\frac{\tilde\mu_\ast}{\sqrt{2}\sigma_\ast}\right)-Q\left(w-\frac{\tilde\mu_\ast}{\sqrt{2}\sigma_\ast}\right)=
\begin{dcases}
-\ln\left(1-\frac{2\tilde\mu_\ast}{\sigma_\ast^2}\right) & x>S_0,\\
\ln\left(1-\frac{2\tilde\mu_\ast}{\sigma_\ast^2}\right) & x<S_0.
\end{dcases}
\label{eq:Q}
\end{equation}
The condition $1-2\tilde\mu_\ast/\sigma_\ast^2>0$ ensures that the term $\ln(1-2\tilde\mu_\ast/\sigma_\ast^2)$ takes a real value.

Let us focus on the $\tilde\mu_\ast>0$ case.
According to the theory of convex functions~\cite{Roberts}, for any strictly convex function $\psi$ and constant $a>0$, the function $\psi(w+a)-\psi(w-a)$ is an increasing function of $w$.
Hence, the left-hand side of Eq.~\eqref{eq:Q} is an increasing function of $w$, and satisfies
\[
\lim_{w\to-\infty}\left[Q\left(w+\frac{\tilde\mu_\ast}{\sqrt{2}\sigma_\ast}\right)-Q\left(w-\frac{\tilde\mu_\ast}{\sqrt{2}\sigma_\ast}\right)\right]=-\infty,
\quad
\lim_{w\to\infty}\left[Q\left(w+\frac{\tilde\mu_\ast}{\sqrt{2}\sigma_\ast}\right)-Q\left(w-\frac{\tilde\mu_\ast}{\sqrt{2}\sigma_\ast}\right)\right]=0.
\]
That is, the left-hand side of Eq.~\eqref{eq:Q} is always negative and can take any negative value by tuning $w$.
Noting that $\ln(1-2\tilde\mu_\ast/\sigma_\ast^2)<0$ for $\tilde\mu_\ast>0$, we conclude that Eq.~\eqref{eq:Q} for $x>S_0$ does not have solution and for $x<S_0$ has a unique solution.
As stated in the $\tilde\mu_\ast=0$ case, the solution $w<0$ corresponds to $x>S_0$, and the peak of $f_\text{uni}(x)$ is at $x=S_0$.

For the $\tilde\mu_\ast<0$ case, the left-hand side of Eq.~\eqref{eq:Q} becomes positive-valued decreasing function and $\ln(1-2\tilde\mu_\ast/\sigma_\ast^2)>0$.
Therefore, Eq.~\eqref{eq:Q} for $x>S_0$ does not have solution and for $x<S_0$ has a unique solution, which indicates that the $f_\text{uni}(x)$ is unimodal.

\section*{Acknowledgments}
The authors are grateful to referees for providing details regarding the double Pareto-lognormal distribution.
One of the authors (K.Y.) was supported by a Grant-in-Aid for Scientific Research (C) 19K03656 and 23K03264 from Japan Society for the Promotion of Science.

\end{document}